\input harvmac
\input epsf
\epsfverbosetrue

%

\def\b0{\bar{0}}
\def\b4{\bar{4}}

\Title{MRI-PHY/P981170}{\vbox{\centerline{
Computation of Lickorish's Three Manifold Invariants }
\vskip 0.1 truein
\centerline{using Chern-Simons Theory}
}}
\smallskip
\centerline{P. Ramadevi$^{(a)}$ \foot{E-mail : 
rama@mri.ernet.in, rama@theory.tifr.res.in} and 
Swatee Naik$^{(b)}$\foot{E-mail: naik@unr.edu, naik@math.tifr.res.in}}
\bigskip
\centerline {$^{(a)}${\it Mehta Research Institute of Mathematics 
and Mathematical Physics,}} 
\centerline{\it Allahabad 211 019, INDIA}
\medskip
\centerline{${(b)}${\it Department of Mathematics, University of Nevada, Reno, NV 89557, USA}}
\centerline{ Currently Visiting: {\it School of Mathematics, 
  Tata Institute of Fundamental Research,}}
\centerline{\it{Mumbai - 400 005, INDIA}} \vskip1cm

It is well known that any three-manifold can be obtained
by surgery on a framed link in $S^3$.
Lickorish gave an elementary proof for the existence of the
three-manifold invariants of Witten using
a framed link description of the manifold
and the formalisation of the bracket polynomial as
the Temperley-Lieb Algebra.
Kaul determined three-manifold invariants
from link polynomials in $SU(2)$ Chern-Simons theory.
Lickorish's formula for the invariant
involves computation of bracket polynomials of several
cables of the link. We describe an easier way of obtaining
the bracket polynomial of a cable using representation theory
of composite braiding in $SU(2)$ Chern-Simons theory.
We prove that the cabling corresponds to taking tensor products of 
fundamental representations
of $SU(2)$. This enables us to verify
that the two apparently distinct
three-manifold invariants are equivalent 
for a specific relation of the polynomial variables.

\Date{January 1999}
\nref\wit{E. Witten, {\it Quantum field theory and Jones polynomials},
Commun. Math. Phys. {\bf 121}, 351-399 (1989)}

\nref\joone{V.F.R.~Jones, {\it A polynomial invariant for knots via von
Neumann algebras}, Bull.~Am. Math.~Soc. {\bf 12} (1985) 103-111.}

\nref\jotwo{V.F.R.~Jones, {\it Hecke algebra representations of braid
groups and link polynomials}, Ann.~of Math. {\bf 126} (1987) 335-388.}

\nref\lione{ W.B.R. Lickorish, {\it A representation of combinatorial
3-manifolds}, Annals of Math.~{\bf 76} (1962) 531-538.}

\nref\wallace{A.D. Wallace {\it Modifications of cobounding manifolds},
Canad.~J.~Math. {\bf 12} (1960) 503-528.}

\nref\ki{R.~Kirby, {\it Calculus of Framed Links in $S^3$},
Invent.~Math.~{\bf 45} (1978) 35-56.}

\nref\wad{M. Wadati, T. Deguchi and Y. Akutsu,
{\it Exactly solvable models and knot theory}, Phys. Rep. {\bf 180},
247-332 (1989)}

\nref\kir{A.N. Kirillov and N.Yu. Reshetikin, {\it Representation algebra
$U_q (sl_2)$, q-orthogonal polynomials and invariants of links},
LOMI preprint E-9-88;see  also in: New Developments in
the Theory of Knots. ed. T. Kohno, World Scientific (Singapore, 1989).}

\nref\RT{N.Yu.. Reshetikin, V.G. Turaev, {\it Invariants of 3-manifolds
via link polynomials and quantum groups}, Invent.~Math. {\bf 103} (1991)
547-597.}

\nref\KM{R.~Kirby, P.~Melvin, {\it The 3-manifold invariants of
Witten and Reshetikin-Turaev for sl$(2, \bf C)$},
Invent.~Math.~{\bf 105}, 473-545 (1991)}

\nref\litwo{ W.B.R. Lickorish, {\it Invariants of 3-manifolds from the
combinatorics of the Jones polynomial}, Pacific J.~Math. {\bf 149} (1991)
337-347.}

\nref\li{ W.B.R. Lickorish, {\it Three Manifolds and Temperley Lieb
Algebra},
Math. Ann. {\bf 290}, 657-670 (1991)}

\nref\lo{W.B.R. Lickorish, {\it Calculations with the Temperley-Lieb
algebra}, Comment Math Helvetici. {\bf 67}, 571-591 (1992)}

\nref\kaufl{L.H. Kauffman and S.L. Lins, Temperley-Lieb Recoupling Theory
and Invariants of 3-Manifolds, Annals of Math.~Studies, no. 134, Princeton
University Press, Princeton, NJ, 1994.}

\nref\thesi{P. Ramadevi, {\it Chern-Simons theory as a theory
of knots and links}, Ph.D. Thesis}

\nref\kau{R.K. Kaul, {\it Chern-Simons Theory, Knot Invariants,
Vertex models and Three-manifold Invariants}, hep-th 9804122}

\nref\trg{R.K. Kaul and T.R. Govindarajan, {\it Three dimensional
Chern-Simons theory as a theory of knots and links}, Nucl. Phys.
{\bf B380}, 293-333 (1992);
{\it Three dimensional Chern-Simons theory as a theory of knots and
links II: multicoloured links}, Nucl. Phys. {\bf B393}, 392-412 (1993)}

\nref\kcmp{R.K. Kaul, {\it Complete solution of $SU(2)$
Chern-Simons theory : Chern-Simons theory, coloured-oriented braids
and link invariants}, Commun. Math. Phys.{\bf 162}, 289-319 (1994)}

\nref\ram{P. Ramadevi, T.R. Govindarajan, R.K. Kaul,
{\it Three dimensional Chern-Simons theory as a theory of knots and
links III: Compact an arbitrary compact semi-simple group},
Nucl. Phys. {\bf B402}, 548-566 (1993)}

\nref\rom{P. Ramadevi, T.R. Govindarajan, R.K. Kaul, {\it Representations
of Composite braids and invariants for mutant knots and links in
Chern-Simons field theories},
Mod. Phys. Lett. {\bf A10}, 1635-1658 (1995)}

\nref\reid{K. Reidemeister, {\it Knot Theory}, BCS Associates Moscow,
Idaho,
USA (translated from Knotentheorie, Ergebnisse der Mathematik und
ihrer Grenzgebiete, (Alte Folge), Band 1, Heft 1, Springer 1932).}

\nref\bir{J. Birman, {\it Braids, Links, and Mapping Class Groups}, Annals
of
Mathematics Studies, No. 82. Princeton University Press, Princeton, N.J.;
University of Tokyo Press, Tokyo, 1974.}

\nref\lang{S. Lang, {\it Algebraic Number Theory}, Springer, New York, 1986.}

\newsec{Introduction}
Classification of three dimensional manifolds has been a long standing 
problem. Witten \wit\ succeeded in giving an intrinsically three-dimensional
definition for Jones' type polynomial invariants \refs{\joone, 
\jotwo} of links using a 
topological quantum field theory known as Chern-Simons theory. 
Witten's approach gives rise to three-manifold invariants $Z(M)$ 
(also called the partition function for the manifold $M$) via surgery on
framed links.

The existing definitions of three-manifold invariants rely on two results:

\noindent
{\bf I}: { \it The fundamental theorem of Lickorish \lione\ and
Wallace \wallace\ that any connected, closed, orientable three-manifold
can be obtained by surgery on a framed link in $S^3$.}

\noindent
{\bf II}: {\it The theorem due to Kirby \ki\ that two framed links determine
the same three-manifold if and only if they are
related by a sequence of diagram moves which are referred to as
Kirby moves.}

\noindent
It follows from Kirby's theorem that
invariants of framed links which are unchanged under Kirby moves
give an invariant of three-manifolds.

Computations of the Witten invariant 
are achieved by exploiting the connection between
Chern-Simons field theory and a two dimensional conformal field theory
known as Wess-Zumino conformal field theory. Though there
are questions about measure in the functional integral formulation
of Chern-Simons theory, 
the computed values of these invariants agree with
the ones obtained from other mathematically
rigorous approaches- viz., exactly solvable two dimensional
statistical mechanical models \wad, quantum groups \refs{\kir- \KM}, 
Temperley-Lieb algebra \refs{\kaufl, \li}. 
The interconnections between Chern-Simons theory,
Wess-Zumino conformal field theory, solvable models and
quantum groups have been summarised in Refs.\refs{\thesi, \kau}.

Given a primitive $4r$th root of unity, Lickorish
\refs{\li, \lo} defines 
an invariant $F_l(M)$
as a linear combination of bracket polynomials of 
cables of a framed link on $S^3$ which under surgery gives the 
three-manifold $M$.
The cabling is necessary for the preservation of the 
invariant under Kirby moves. 
However, it introduces a large number of crossings in the
diagram, and determining the bracket polynomials of link diagrams with
several crossings is extremely cumbersome by the recursive method.
Hence, the computation of Lickorish's invariant is not easy.

Witten showed  that the Jones' and
HOMFLY polynomials of links in $S^3$ correspond to expectation 
values of Wilson loops carrying defining representation of the $SU(2)$ and 
$SU(N)$ gauge groups respectively \wit. 
This method has been generalised to arbitrary
higher dimensional representations of any compact semi-simple
group resulting in  a whole lot of new invariants of (framed) links 
in $S^3$ \refs{\thesi-\ram}. 
We refer to these field theoretic invariants
as generalised invariants. 
Unlike Jones', HOMFLY and bracket polynomials, the
generalised invariants cannot be solved completely by the recursive
method. 
Hence, a direct method of evaluating these
was developed in Refs. \refs{\thesi-\ram}. 

By construction
these generalised invariants depend on the framing chosen for
the link. 
However, by fixing
the framing to be standard, i.e., one in which 
the linking number of the link with its frame is zero, 
ambient isotopy invariants of links are obtained.
In Refs. \refs{\thesi -\ram}, the emphasis was 
on obtaining ambient isotopy invariants of links
and hence computations were done in standard framing.
In the present problem, we require a field theoretic 
presentation for the bracket polynomial
of a link diagram. Bracket polynomials
are regular isotopy invariants. 
So we choose the vertical or the black-board framing.
We show how the Chern-Simons invariant $P_{1,1, \ldots ,1}[D_L] (q)$
in vertical framing for defining representation placed on any  
$n$-component link $L$ is related to 
the bracket polynomial $\langle D_L \rangle (A)$
provided the polynomial variables satisfy
\eqn\zer{
q^{1/4}=-A .~}
The relation is proved by first establishing a connection  
between the field theoretic
invariant $P_{1,1,\ldots ,1}[D_L](q)$ and the Jones' polynomial,
and then using the well-known relationship between the Jones' polynomial
and the bracket polynomial (See Section 3 and Theorem 1).

Using the generalised regular isotopy invariants 
in $SU(2)$ Chern-Simons theory, Kaul \kau\ has derived a three-manifold 
invariant $F_k[M]$. 
It is mentioned in \kau\ that wherever computed, 
the Chern-Simons
partition function Z[M] turns out to be the same as his three-manifold
invariant except for the normalisation. That is, $ F_k(M) = Z[M] / Z[S^3]$.
We expect Kaul's invariant to be equivalent to Witten's invariant 
for an arbitrary $M$. We show the equivalence by the following 
sequence of steps:

\noindent
{\it (i) We prove that the Lickorish's and Kaul's invariants 
are equivalent 
with the polynomial variables obeying \zer\
by finding an elegant 
and easier method of determining bracket polynomials of cables 
of link diagrams. (See Theorem 2.) This is achieved by using the 
techniques developed in Ref.\rom\ -viz., representation theory 
of composite braids. 

\noindent
\it (ii) In \lo\ the equivalence between Lickorish's invariant and the 
Reshetikin-Turaev invariant was established. The Reshetikin-Turaev
invariant is considered as a reformulation of Witten's invariant using
quantum groups (see \RT, \KM).} 

\noindent
It follows that 
Kaul's invariant is a reformulation of the 
Reshetikin-Turaev invariant
in terms of the Chern-Simons generalised framed link invariants.

The plan of the paper is as follows.
In Section 2, we describe Lickorish's three-manifold
invariant obtained using bracket polynomials.
We present in Section 3, the techniques used in evaluating Chern-Simons 
regular isotopy invariants. 
We show detailed computations for the Hopf link and then
generalise the method to prove Theorem 1.
Then, we define the three-manifold 
invariant derived by Kaul from these generalised 
link invariants. In Section 4, we study the representation
theory of parallel copies of braids.  This is essential 
to compute directly the bracket polynomial 
for cables of link diagrams
without going through the 
extremely tedious process of recursive
evaluation. We show the details of our computation for the 
$(2,3)$-cable of the Hopf link and generalise the techniques 
to arbitrary links.
In the concluding section, we show that
the three-manifold invariants obtained by 
Lickorish's approach and the field theory approach
are equivalent and thereby provide an
easier method for computing Lickorish's invariant.

\newsec{Lickorish's Three Manifold Invariant}
We briefly present the salient features of Lickorish's
three-manifold invariant obtained from bracket polynomials.

An $n$-component link $L$ in $S^3$ is a subset of $S^3$ homeomorphic
to the union of $n$ disjoint circles. 
A framing ${\bf f} = (f_1, f_2, \ldots , f_n)$ on $L$ is an assignment of an integer to each
component of $L$. 
A regular projection of a link in the plane is one with transverse
double points as the only self-intersections. These double points
are referred to as crossings. A link diagram is 
obtained from a regular projection by marking the under crossing arc
at a crossing with a break to indicate that part of the curve dips
below the plane. Regular isotopy is an equivalence relation on 
the set of link diagrams. It is generated by Reidemeister moves II and III
\reid. A framed link $[L,{\bf f}]$ can be represented by a 
diagram $D_L$ in the plane such that 
the framing on each component of $L$ equals the sum of crossing
signs in the part of the diagram that represents that component.
Given a link diagram, the framing it represents will be called the
blackboard framing or the vertical framing.

The bracket polynomial as normalised in \li\ is a function
\eqn\libra{
\langle ~ \rangle \colon \{ {\rm link\ diagrams\ in\ }({\bf R}^2 \cup \infty )
{\rm \ of \ unoriented \ links\ } \} \to {\bf Z} [ A^{\pm 1} ]} 
defined by the following three properties.

(i) $\langle \phi \rangle =1;$

(ii) $\langle D_L \cup U \rangle = (-A^2 -A^{-2}) <D_L>,$  where $U$ 
is a component with no crossings;

(iii) $\langle
{\epsfxsize=2mm \epsfbox{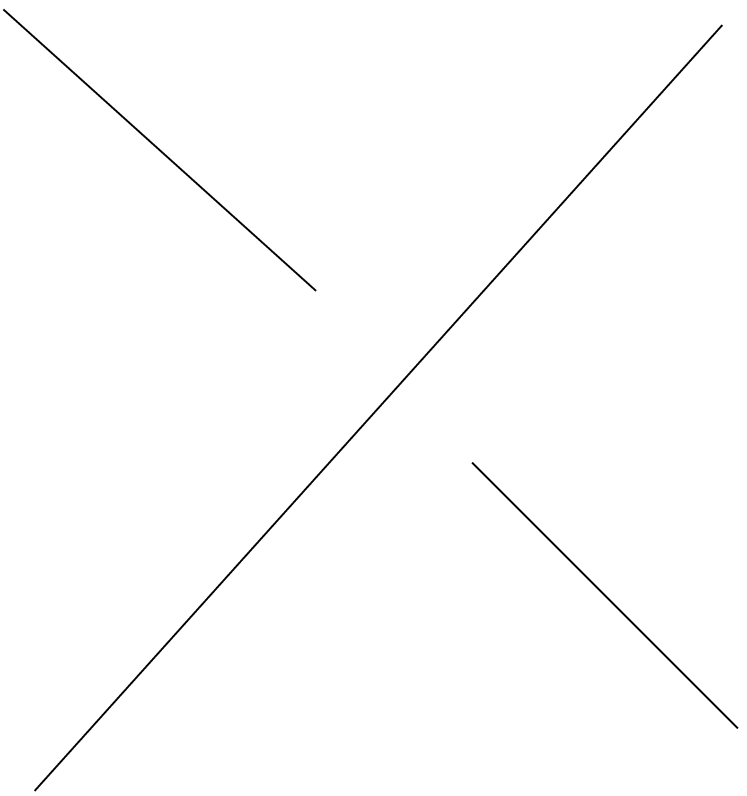}} \rangle = A \langle
{\epsfxsize=2mm \epsfbox{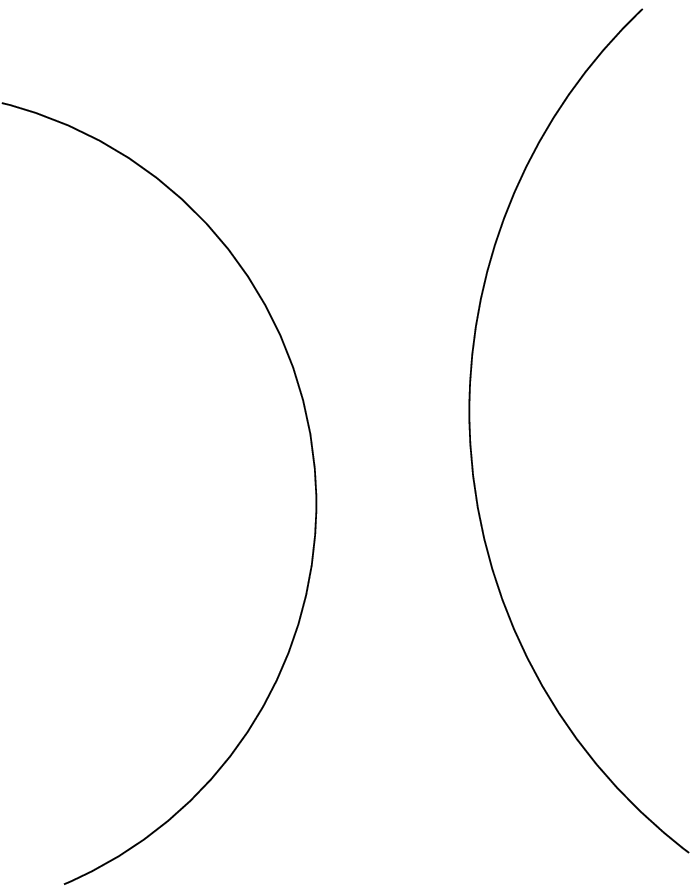}} \rangle + A^{-1} \langle 
{\epsfxsize=2mm \epsfbox{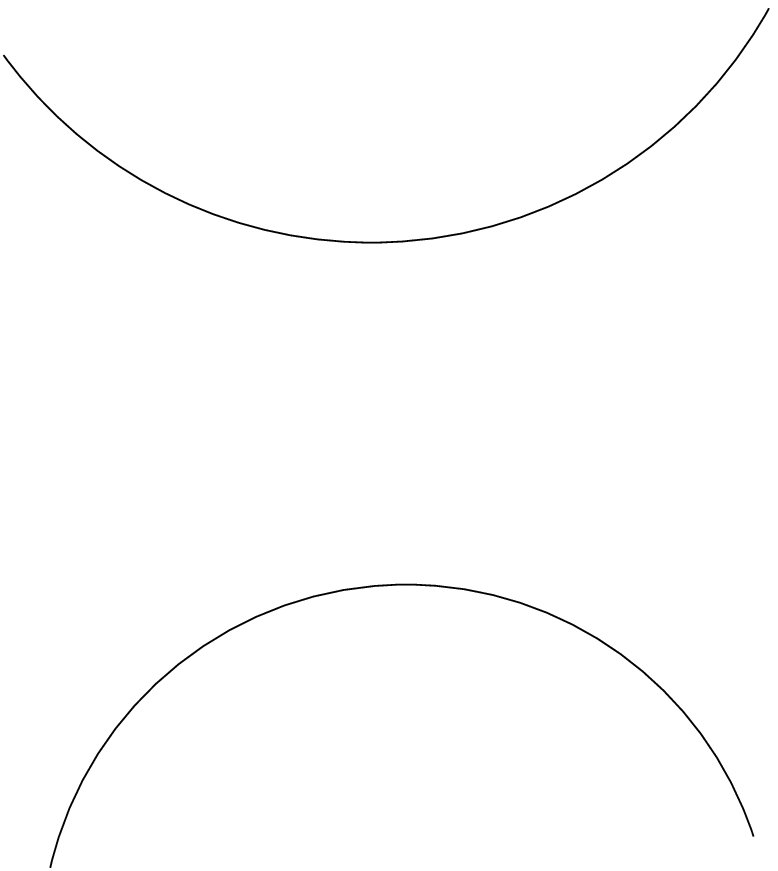}} \rangle $ 
where 
this refers to three diagrams identical 
except at one crossing where they look as shown.

This is a regular isotopy invariant of link diagrams.

In order to define a three-manifold invariant using the bracket,
Lickorish obtains an expression which is invariant under Kirby
moves on link diagrams. Before we can state Lickorish's result we need the
following definition.
\vskip5mm

\noindent{\bf Definition}: For a diagram $D_L$ representing an
$n$ component framed link $[L,{\bf f}]$, and a given $n$-tuple of nonnegative
integers ${\bf c} = [c(1), c(2), \ldots ,c(n)]$,
a ${\bf c}$-cable ${ \bf c} * D_L$ is defined as 
the diagram obtained by replacing the $i$th component of $L$ in $D_L$
by $c(i)$ copies all parallel in the plane.
\vskip5mm

As in \lo, let $r$ be a fixed integer, $r \ge 3$, and let 
$C(n,r)$ denote the set
of all functions ${\bf c} \colon \{1,2, \ldots , n \} \to 
\{ 0, 1, \ldots , r-2 \} $.
Let $A$ be a primitive $4r$-th root of unity.
Let $G = G(A)$ be the Gauss sum $\sum_{l=1}^{4r} A^{l^2}$
and let $\bar G$ denote the complex conjugate of $G$. 

\vskip5mm

\noindent{\bf Proposition} \li\ \lo\ :
Let $M$ be a three-manifold obtained from $S^3$ by surgery 
on an $n$-component framed link 
represented by a diagram $D_L$, and let
$\sigma$ and $\nu$ be the 
signature and the nullity of the linking matrix, respectively. 
Then 
\eqn\one{F_l (M)=\kappa^{\sigma+ \nu - n \over 2} \sum_{{\bf c} \in C(n,r)}
\lambda_{c(1)} \lambda_{c(2)} \ldots ,\lambda_{c(n)} 
\langle {\bf c} * D_L \rangle }
is an invariant of the three-manifold
with $\kappa$ and $\lambda_c$ given by
\eqn\likap{\kappa = (-1)^{r+1} A^6 (\bar G /G),} 
\eqn\lilam{\lambda_c = 2G^{-1} A^{r^2+3} {\sum_j}_{0 \le 2j \le r-2-c}
(-1)^{c+j} \left( \matrix{c+j \cr j} \right) \left( A^{2(c+2j+1)}-A^{-2(c+2j+1)}
\right) . }
\vskip5mm

In \li\ only the existence and uniqueness  
of the $\lambda _c$ was shown without giving any method of
computation. The formulas \likap\ and \lilam\ for $\kappa$
and $\lambda _c, \ 0 \le c \le n,$ were obtained in \lo. 
The equivalence of
Lickorish invariant to Reshetikin-Turaev invariant
as in the Kirby-Melvin \KM\ formulation is established in \lo\
for  $A = -e^{\pi i \over 2r}$. 

In spite of having these formulas the computation of the invariant 
$F_l$ is quite difficult as one has to compute 
bracket polynomials of the {\bf c}-cables which is
very cumbersome. As mentioned in \li\ if $r=6$, and $D_L$ is a standard
diagram of the trefoil with only $3$ crossings, the computation of $<4*D_L>$
using the definition of the bracket as given above would
involve $2^{48}$ operations. 
In the next two Sections, we will concentrate on the link
invariants from Chern-Simons field theory with the motivation
of finding an easier method of computing the bracket
polynomials of cables. 

\newsec{$SU(2)$ Chern-Simons Theory and Link Invariants}

Now, we briefly present the methods employed in the direct evaluation
of Chern-Simons link invariants \refs{\wit ,\thesi-\ram}. 

The metric independent action $S$ of the $SU(2)$ Chern-Simons theory
on any three-manifold $M^3$ is  given by
\eqn\csac{ S = {k \over 4 \pi} \int_{M^3} A \wedge dA + {2 \over 3} 
A \wedge A \wedge A}
where $A$ is a one-form (matrix-valued in the Lie algebra
of a compact semi-simple Lie Group $SU(2)$). Explicitly, 
$A= A_{\mu}^a T^a dx^{\mu}$ where $T^a$ are the
$SU(2)$ generators and $k$ is the coupling constant. 

The Wilson loop operators of any link $L$ embedded in $M^3$ 
is given by
\eqn\wil{ W_{R_1 R_2 \dots R_n}(L)= \prod_i W_{R_i} (C_i) = 
\prod_i \{Tr_{R_i}P \exp\oint_{C_i} A_{\mu} dx^{\mu} \}}
where $C_i$'s are the component knots of the link $L$
and $Tr_{R_i}$ refers to the trace over the $SU(2)$ representation
$R_i$ placed on  the component $C_i$. 

The link invariants are given by the expectation
value of the Wilson loop operator:
\eqn\lin{ P_{2R_1,2R_2, \ldots , 2R_n}[D_L]=\langle W_{R_1,R_2, \ldots , R_n} (L) \rangle = 
{\int [{\cal D}A_{\mu}]~ \prod_i W_{R_i} (C_i)~ e^{i S} \over
\int [{\cal D}A_{\mu}]~  e^{i S}}~,}
where $D_L$ denotes a diagram representing the framed
link $L$ in vertical or black-board frame.
This functional integral over the space of matrix valued one forms $A$ 
is evaluated by exploiting the connection between the 
Chern-Simons theory in three dimensional space with boundary
and the $SU(2)_k$ Wess-Zumino conformal field theory on the 
two dimensional boundary \wit. The computation
of these invariants has been considered in detail in  
Refs. \refs{\thesi -\ram}. 
We summarize this below.

We represent a link as the closure of a braid.
The braid group ${\cal B}_m$ consisting of $m$-strand braids
is generated by $b_i, 1 \leq i \leq m-1$, where $b_i$ represents 
a right-handed half-twist between the $i$-th strand and
the $i+1$-st strand. The inverse 
$b_i^{-1}$ corresponds to a left-handed half-twist.
For a standard reference on braid groups see \bir.
In order to illustrate the technique of direct
evaluation of link invariant \lin,
we will take the example of the Hopf link.
The details we present in this example are general 
enough for deriving invariants of links obtained from 
a four-strand braid either by platting or capping.
In Theorem 1 we generalise the technique to arbitrary links.

Consider the Hopf link $H$, obtained from a
four-strand braid (drawn as a closure of a two-strand braid), 
embedded in $S^3$ as shown in Fig.1. 
Let $j_1$ and $j_2$
denote the representations placed on the two component knots of $H$.

\centerline{\epsfxsize=3in \epsfbox{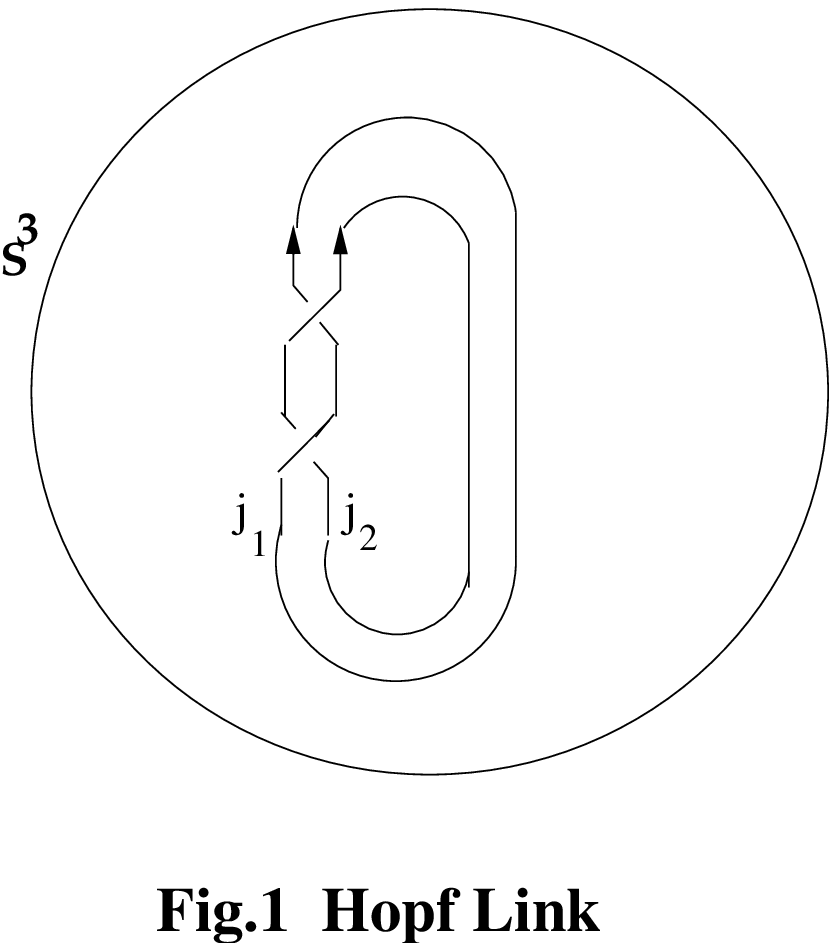}} 
Let us slice the three-manifold $S^3$ into two 
three dimensional balls as shown in 
Fig.2(a) and (b). 
The two dimensional $S^2$ boundaries 
of the three-balls are oppositely oriented and have four 
points of intersections with the braid, which we refer to as
four punctures.

\centerline{\epsfxsize=3in \epsfbox{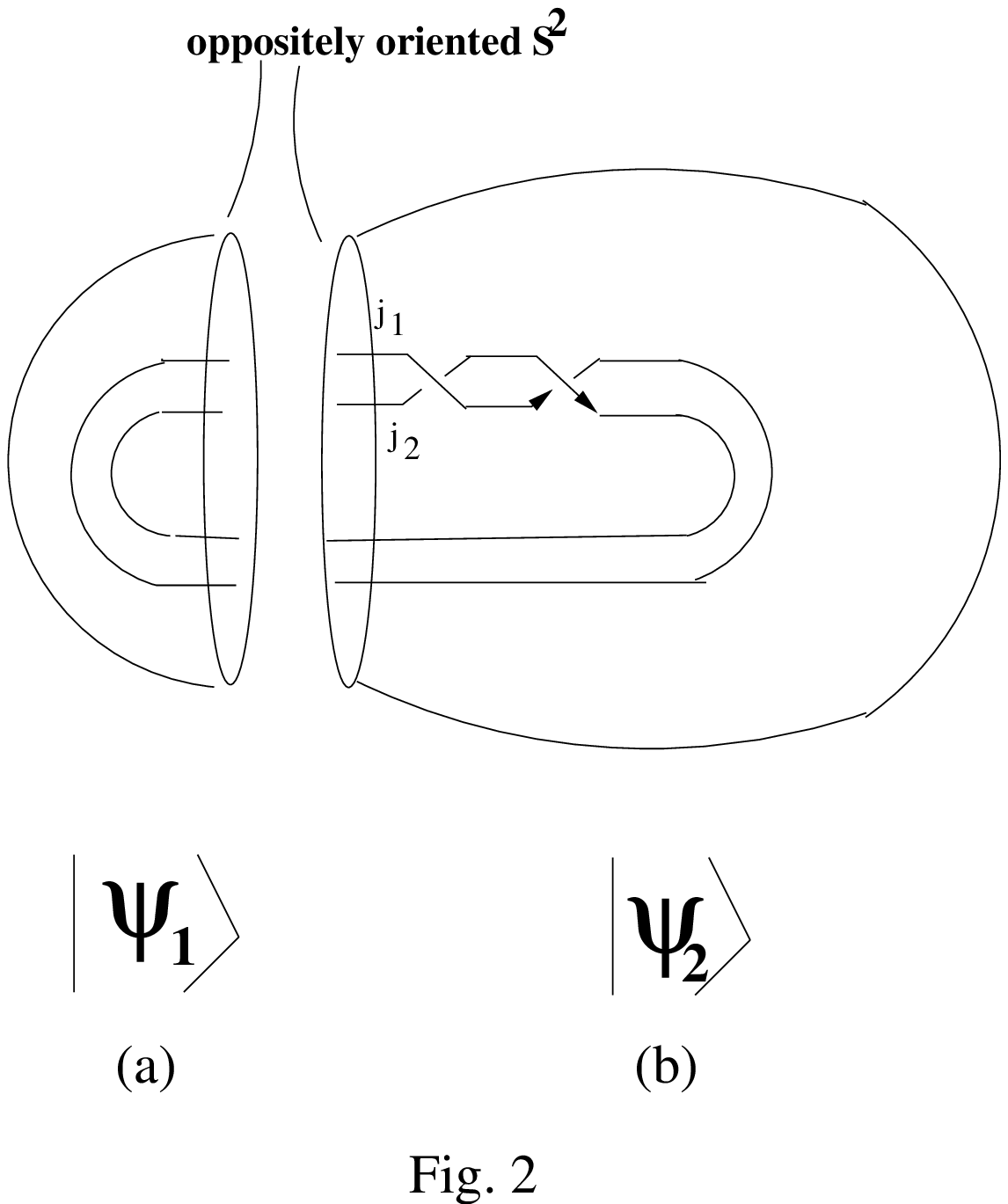}}

Now, exploiting the connection between 
Chern-Simons theory and Wess-Zumino conformal field theory,
the functional integrals of these three-balls  correspond to
states in the space of four point correlator conformal blocks
of the Wess-Zumino conformal field theory \wit. The dimensionality of 
this space is dependent on the representation of $SU(2)$ placed on
the strands and the number of punctures on the boundary. 
In the present example, the dimension of the space is 
\eqn\dim{
{\rm min} (2j_1 +1 , 2j_2 +1, k-2j_1+1,k-2j_2 +1)}
These states can be written in a suitable basis. 
Two such choices of bases ($|\phi_s^{side} \rangle )$, and 
$(|\phi_t ^{cent} \rangle$) are pictorially depicted in 
Fig.3(a),(b). Here $s \in j_1 \otimes j_2$ and $t \in {\rm min}
(j_1 \otimes j_1, j_2 \otimes j_2)$ where $\otimes$ 
(also called tensor product notation) is defined as:
\eqn\dsu{ j_1 \otimes j_2 = |j_1-j_2| \oplus |j_1 -j_2|+1 \oplus \ldots, 
\oplus {\rm min}(k-j_1-j_2, j_1+j_2)}
with $\oplus$ usually referred to as  direct sum.

\centerline{\epsfxsize=3in \epsfbox{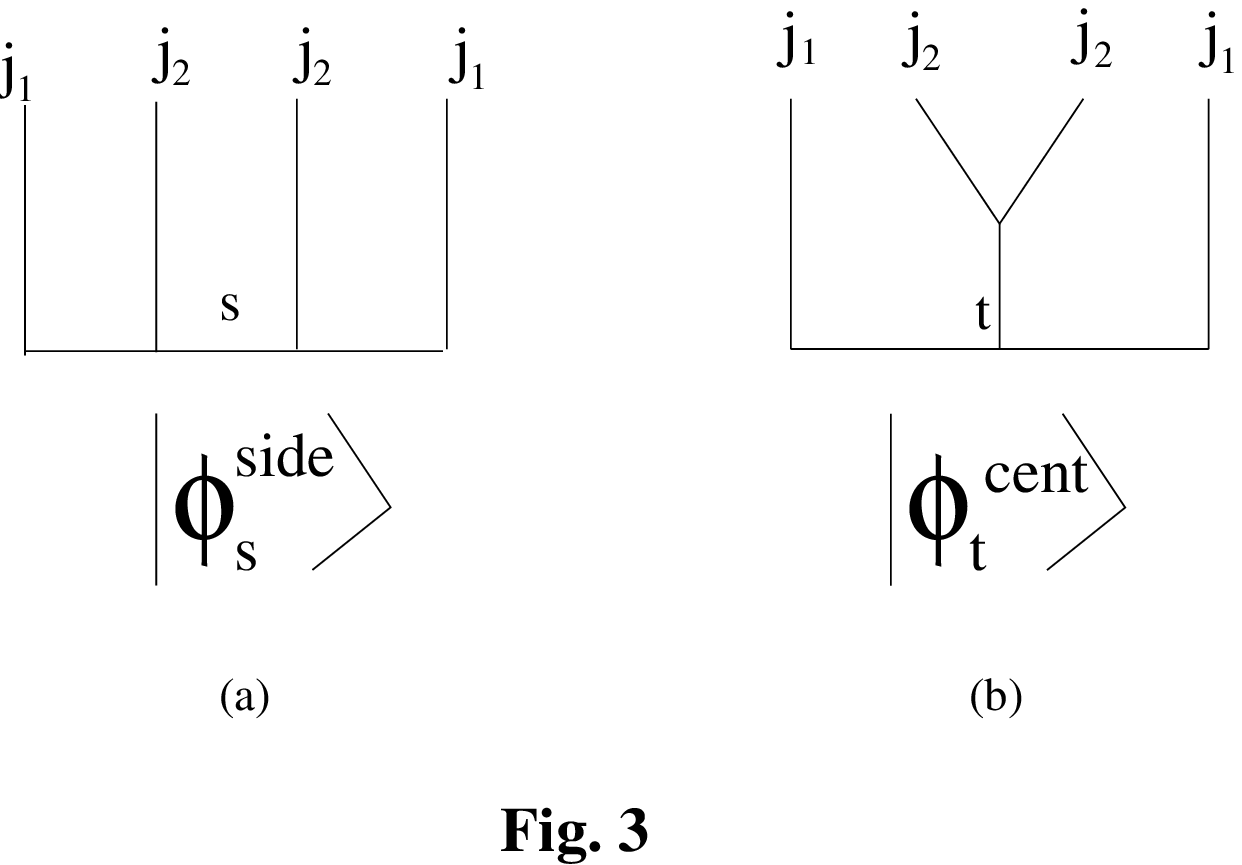}}
The basis $|\phi_s^{side} \rangle$ is chosen when the braiding is done
in the side two parallel strands. In other words, it is the 
eigen basis corresponding to the  
generators $b_1$ and $b_3$:
\eqn\eigen{b_1^2 |\phi_s^{side}\rangle =
b_3^2 |\phi_s^{side}\rangle =
(\lambda_{s,R}(j_1,j_2)^{(+)})^2 
|\phi_s^{side}\rangle}
with the eigenvalues $\lambda_s ^{(+)}(j_1,j_2)$ for the left-handed and 
right-handed half-twists in parallel strands (vertical framing) being:
\eqn\pevalue{\lambda_{s,R}^{(+)}(j_1,j_2) =
\left(\lambda_{s,L}^{(+)}(j_1, j_2)\right)^{-1}= 
(-1)^{j_1+j_2-l} q^{[j_1(j_1+1)+j_2(j_2+1)-s(s+1)]/2}.}
Similarly for braiding in the middle two anti-parallel strands $b_2$,
we choose the basis $|\phi_t^{cent} \rangle$:
\eqn\eign{b_2^2 |\phi_t^{cent}\rangle = (\lambda_{t,R}(j_2,j_2)^{(-)})^2 
|\phi_t^{cent}\rangle}
with the eigenvalues in anti-parallel strands being:
\eqn\apeval{
\left(\lambda_{t,R}^{(-)}(j_1, j_2)\right)^{-1}=
\lambda_{t,L}^{(-)}(j_1,j_2) =
(-1)^{|j_1-j_2|-l} q^{[j_1(j_1+1)+j_2(j_2+1)-t(t+1)]/2~.}}

These two bases are related by a duality matrix 
$$\left( a_{st}\left(\matrix{j_1 &
j_2 \cr j_2 & j_1 \cr}\right) \right), \ 
s \in j_1 \otimes j_2 ,\ t \in {\rm min}
(j_1 \otimes j_1, j_2 \otimes j_2),$$
 defined as:
\eqn\dual{ |\phi_s^{side} \rangle = a_{st}\left(\matrix{j_1 &
j_2 \cr j_2 & j_1 \cr}\right) |\phi_t^{cent} \rangle}
The matrix elements of the duality matrix are the $SU(2)$ quantum
Racah coefficients which are known \refs{\kcmp, \kir}.

For example, the duality matrix for $j_1=j_2={1 \over 2}$ (defining/
fundamental representation of $SU(2)$) is a
$2 \times 2$ matrix:
\eqn\dual{ {1 \over [2]}\left( \matrix{-1 & \sqrt{[3]} \cr
          \sqrt{[3]} & 1}\right)}
where the number in square bracket refers to the quantum number defined as
\eqn\quan{[n] = {q^{n \over 2} - q^{-{n \over 2}} \over 
q^{1 \over 2} - q^{-{1 \over 2}}}}
with $q$ (also called deformation parameter in quantum
algebra $SU(2)_q$) related to the coupling constant $k$ as
$q = \exp({2 i \pi \over k+2})$.
We will see that the invariants are polynomials in $q$.

Now, let us determine the states corresponding to Figs.2(a) and (b).
Since the braiding is in the side two parallel strands, 
it is preferable to use $|\phi_s^{(side)} \rangle$ as basis states. 
Let $|\Psi_1 \rangle$ be the state corresponding to Fig. 2(a). 
Clearly, we can write the state for Fig. 2(b) as
\eqn\brai{|\Psi_2 \rangle = b_1^2 |\Psi_1 \rangle}
This state should be in the dual space as its $S^2$ boundary
is oppositely oriented compared to the boundary in Fig. 2(a).
Then the link invariant is given by 
\eqn\inv{P_{2j_1, 2j_2} [D_H] ~=~\langle \Psi_1 |b_1^2 |\Psi_1 \rangle}
For determining the polynomial, we will have to express 
the states as linear combination of the basis 
states $|\phi_s^{side} \rangle$. The coefficients 
in the linear combination are chosen such that
$$\langle \Psi_1 | \Psi_1 \rangle = P_{2j_1,2j_2}[D_{U^2}]= 
[2j_1 +1] [2j_2 +1]~,$$ 
where $P_{2j_1,2j_2}[D_{U^2}]$ gives the polynomial of the unlink with
2 components. 
\foot{\it We work in the unknot polynomial 
normalisation  $P_{2j}[D_U]=[2j+1]$ with the representation
$j$ placed on unknot. The square bracket denotes the quantum number \quan.} 

The above mentioned restrictions determine the state $|\Psi_1\rangle$ 
(see \trg) as: 
\eqn\sta{|\Psi_1\rangle 
 = \sum_{s=|j_1-j_2|}^{{\rm min}(j_1+j_2, k-j_1-j_2)} \sqrt{[2s+1]}
|\phi_s^{side~} \rangle}
Substituting it in eqn. \inv\ and using the braiding eigenvalue \pevalue,
we obtain
\eqn\poly{P_{2j_1,2j_2}[D_H]~=~\sum_{s=|j_1-j_2|}^{{\rm min}
(j_1+j_2, k-j_1-j_2)} [2s+1]
 (\lambda_{s,R}^{(+)}(j_1,j_2))^2~.} 
For $j_1=j_2= 1/2$, we get the following polynomial.
\eqn\poyy{P_{1,1}[D_H]~=~q^{3\over 2} + (q + 1 + q^{-1}) q^{-1\over 2}~=~
                       q^{3\over 2} + q^{1\over 2} + q^{-1 \over 2} 
                        + q^{-3 \over 2}.}
The bracket polynomial for Hopf link, represented
by diagram $D_H$,  obtained by the recursive method is
\eqn\poy{\langle D_H \rangle ~=~A^6 + A^{2} +A^{-2} + A^{-6}
                             ~=~P_{1,1}[D_H]  ~\vert_{q^{1/4}=-A}.}

The bracket polynomial $\langle D_L \rangle (A)$
and Jones' polynomial
$V[L](q)$ 
for any $n$-component link 
are related as given below
(see, for instance, \lo\ which uses a different
normalisation\foot{ We work in the unknot normalisation 
:
$V[U]=(q^{1 \over 2}+q^{-{1 \over 2}})$,
$\langle D_U \rangle=-(A^2+A^{-2})$.})
\eqn\jon{
(-1)^n V [L] (q) ~\vert_{q^{1/4}=-A}  = (-A)^{3 \omega}
\langle D_L \rangle (A)~,
}
where $\omega$ is the writhe or the sum of crossing 
signs\foot{In standard literature $\omega$ 
in the exponent may appear with a negative sign. 
This is a matter of replacing the variable $q$ with $q^{-1}$.}
in the diagram $D_L$. 

The Jones' polynomial is
obtained by placing the defining representation $j_1=j_2={1 \over 2}$
on the component knots in standard framing. 
The  standard framing braiding eigenvalue for a right-handed
half-twist in  parallel strands
$\tilde \lambda_{r,R}^{(+)}(j_1,j_2)$ 
is related to the corresponding vertical framing eigenvalue as
\eqn\rela{\tilde{\lambda}_{r, R}^{(+)} ({1 \over 2}, {1 \over 2})~=~ 
q^{3 \over 4} \lambda_{r,R}^{(+)} ({1 \over 2}, {1 \over 2}).}
This equation 
determines the frame correction
factor between the ambient isotopy invariant and the regular isotopy invariant.
Using \poly, it
is clear that for the Hopf link 
the Jones' polynomial $V[H]$ is related to
the invariant in vertical framing as
\eqn\jver{V[H]~=~ ~\sum_{s=0}^{{\rm min}
(1, k-1)} [2s+1] (\tilde{\lambda}_{s,R}^{(+)}(j_1,j_2))^2~ 
~=~q^{3/2} P_{1,1}[D_H].} 
For the diagram in Figure 1, $\omega = 2$, and the number of
components $n=2$. Combining \jon\ and \jver\ we get
${ P_{1,1}[D_H]~\vert_{q^{1 \over 4}=-A}= 
\langle D_H \rangle}$
which confirms eqn. \poy. 

The method elaborated above for a specific example can be generalised for 
any $n$-component link obtained as a closure of an $m$-strand
braid ($n \leq m$). We will briefly outline the steps below.

\vskip5mm

{\bf Theorem 1}: {\it For a diagram $D_L$ of an $n$-component link 
the bracket polynomial and the invariant in vertical
framing are related as 
\eqn\bre{ \langle D_{L} \rangle~\vert_{\left( A =- q^{1 \over 4} \right)} ~=~ 
(-1)^n P_{1,1,\ldots ,1}[D_{L}]  .}}

{\bf Proof}:

\centerline{\epsfxsize=3in \epsfbox{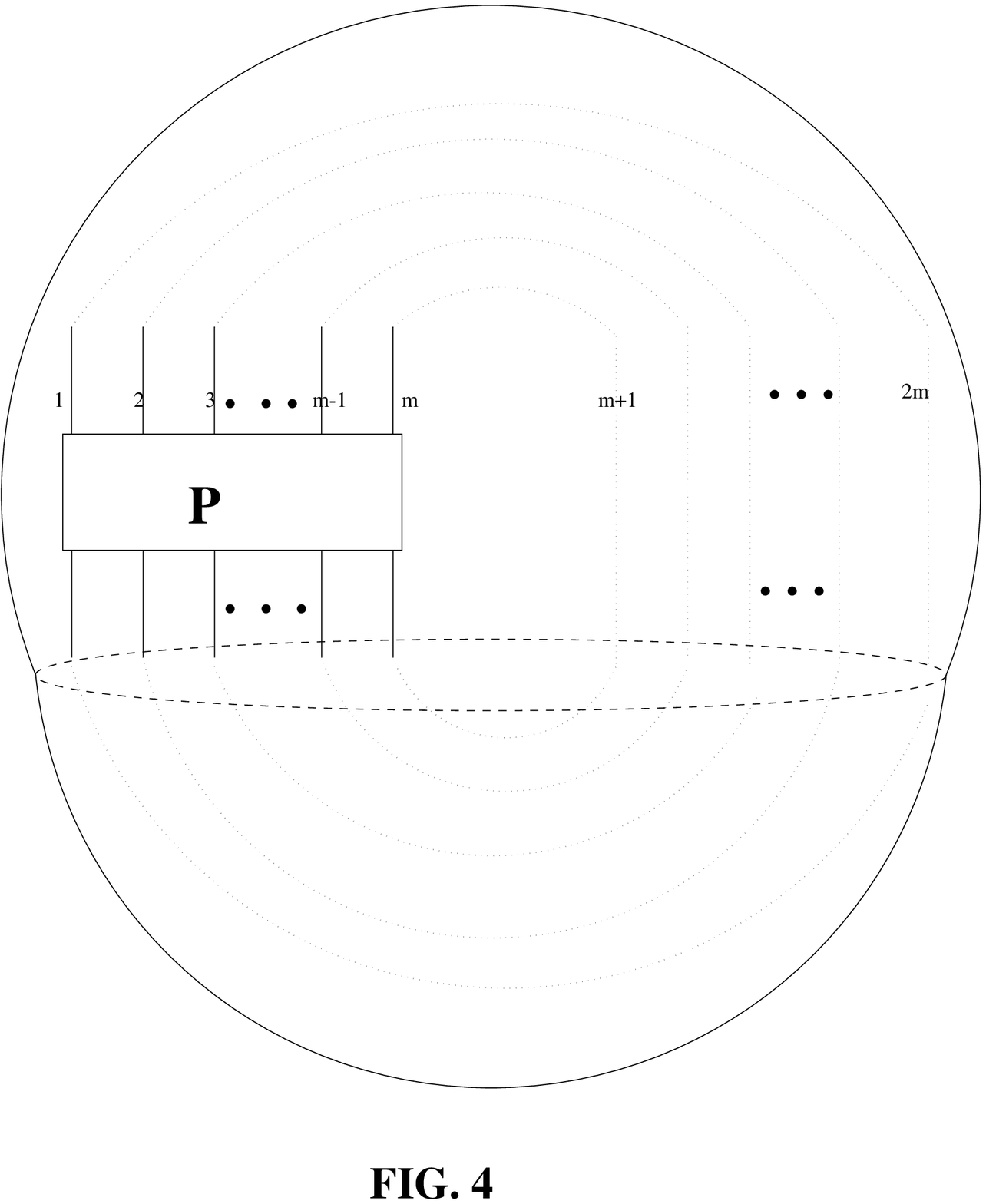}}

Let us  take an arbitrary braid word in the braid group
${\cal B}_{2m}$ denoted by the black box P as shown in
Fig. 4. Here, the dotted lines denote the closure and
the dashed line represents slicing of $S^3$ into
two pieces with oppositely oriented $S^2$ boundaries.
Note that the last $m$ strands are trivial, so this can 
also be considered the closure of an $m$-braid.
\centerline{\epsfxsize=3in \epsfbox{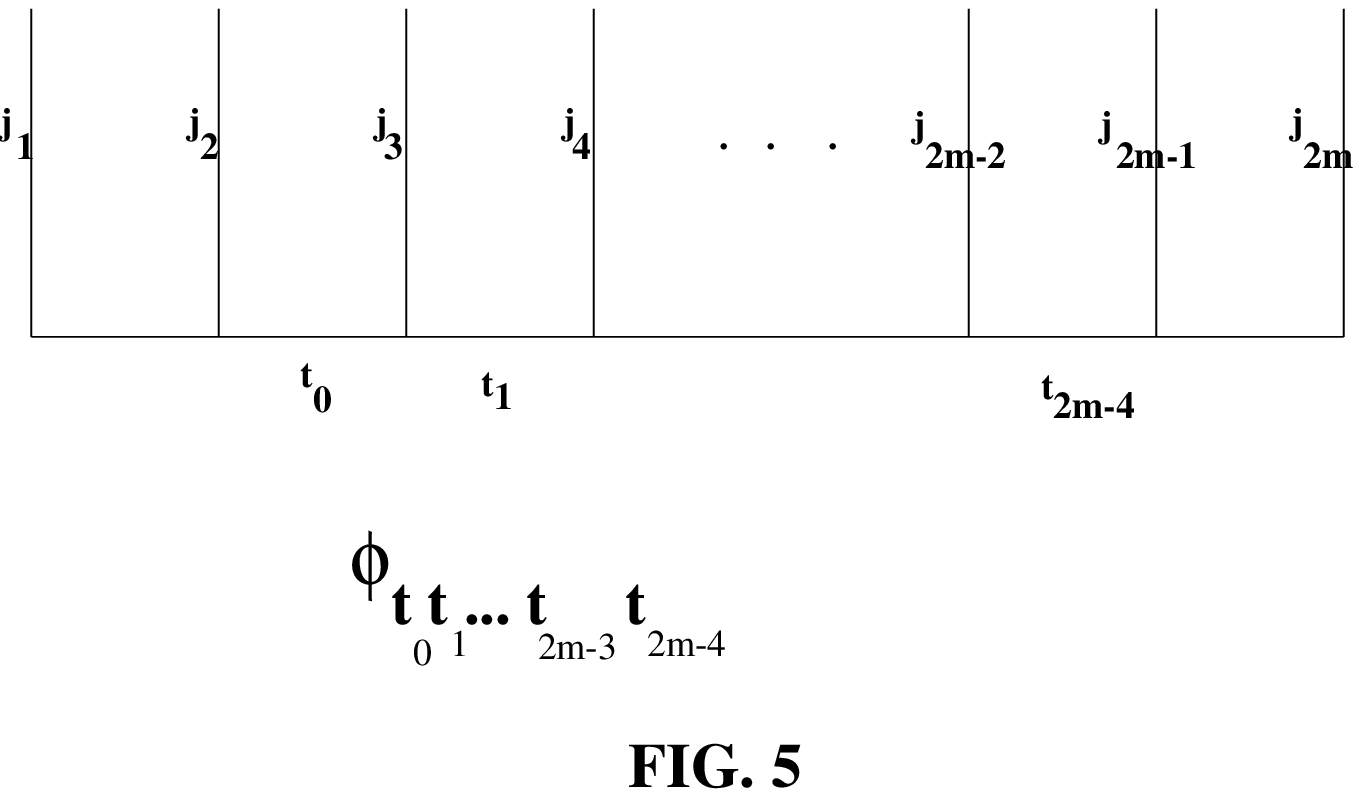}}

The states on the $2m$ punctured surface corresponding
to the two three-balls can be expanded in a suitable basis(see Fig.5):
\eqn\gene{\eqalign{|\Psi_1 \rangle =&
\sum_{t_0,t_1 \ldots ,t_{2m-4}} A_{j_1 \ldots ,j_{2m},t_0,t_1 \ldots ,t_{2m-4}}[P]~
|\phi_{t_0,t_1,\ldots ,t_{2m-4}} \rangle\cr
\langle \Psi_2 | =&
\sum_{t_0,t_1 \ldots ,t_{2m-4}} 
B_{j_1,j_2 \ldots ,j_{2m},t_0,t_1 \ldots ,t_{2m-4}}~
\langle \phi_{t_0,t_1,\ldots ,t_{2m-4}}|}}
where the summation variables $t_i \in t_{i-1} \otimes j_{i+2}$
\dsu\ with $t_{-1}=j_1$ and $t_{2m-3}=j_{2m}$.
The element $A$ in the summation depends on the braid word $P$
and $B$ is chosen such that $\langle \Psi_2 | \Psi_2 \rangle$
gives $\prod_{i=1}^m [2j_i +1]$.

There are various ways to obtain a link by closing a braid. 
Some such ways are discussed in \kcmp.
The closure as shown in Fig.4 will demand that
$$\left( j_{m+1}, j_{m+2} \ldots ,j_{2m} \right)~=~
 \left( j_m, j_{m-1} \ldots ,j_1 \right).$$

Additional restrictions on 
$j_i$ are dictated by the braid word $P$ and the $n$-component link.
The link invariant 
is well-defined only if all strands which correspond to the
same component of the link are marked with the same $j_i$.
(i.e.), at most $n$ of the $j_i$'s can be different.

With these inputs, any $n$-component link invariant
will be
\eqn\nlink{P_{2j_1,2j_2, \ldots ,2j_n} = \langle \Psi_2|\Psi_1 \rangle}

In order to compare this invariant with the Jones' polynomial, we choose
$j_1=j_2= \ldots = j_n=1/2$ and 
as a consequence of the frame correction factor \rela\ between ambient
isotopy (standard framing) and regular isotopy (vertical framing) we have 
\eqn\gver{V [L]~=~ (q^{3 \over 4})^{\omega}
P_{1,1, \ldots ,1}[D_{L}].}
Hence from eqn.\jon\ we have the result \bre\ ({\bf Proved}).
\vskip5mm

The generalised link polynomials \nlink\ of some of the knots up to 
eight crossings and two component links have been tabulated in
Appendix II of \kcmp. 
We have to use vertical framing braiding
eigenvalues \pevalue, \apeval\ instead of the standard framing 
eigenvalues.

With these regular isotopy polynomial invariants, a three-manifold
invariant which respects Kirby's theorem has been
constructed in \kau; the formula being
\eqn\two{F_k (M)=\alpha^{-\sigma[L]} \sum_{{\bf c} \in C(n,k+2)} 
\mu_{c(1)} \mu_{c(2)} \ldots ,\mu_{c(n)} P_{c(1),c(2) \ldots ,c(n)}[D_L],}
where 
\eqn\twoa{\mu_c = {1 \over 2i} \sqrt{2 \over k+2} 
\left( q^{c+1 \over 2}- q^{-(c+1) \over 2} \right), 
~\alpha= (q^{1 \over 4})^{3 k \over 2},} 
and $\sigma$ denotes the signature of the linking matrix 
in a framed link representation of the manifold $M$.

It is not a priori clear whether this formula gives the
same invariant as the Lickorish invariant \one\ 
\foot{\it The presence or absence of the nullity $\nu$
of the linking matrix for the framed link appears to be a 
matter of normalisation as $\nu$ is unchanged under both the
Kirby moves.} with the polynomial variables related as in eqn. \zer.
In order to verify the equality, we need to find a method
of determining brackets of cables in terms of the invariant 
$P_{c(1),c(2), \ldots ,c(n)}$ \nlink.

In the next section, we will present the representation
theory of composite braiding which will prove useful to
directly compute the invariants of 
cables of link diagrams.

\newsec{Composite Braiding and {\bf c}-cable Link invariants}

The representation theory of composite braiding 
involves determination of braiding eigenvalues,
eigen basis and the duality matrix. 
One such representation in standard framing was presented in \ram\ 
in an attempt to distinguish a class of knots called mutants.
In this paper, we have a slightly different composite braiding.
The braiding eigenvalues and eigen basis are derived
in a similar fashion as in Ref. \ram.

{\bf Definition}: A $c$-Composite 
of a given 
braid is obtained by replacing every strand by $c$-strands
and the 
generator $b_i$ by a composite braiding $B_i^{(r)}$
\eqn\com{B_i^{(c)}= b(ci,ci+c-1) b(ci-1,ci+r-2) \dots b(ci-c+1,ci)}
where $b(i,j)=b_i b_{i+1} \ldots ,b_j$. 

For convenience we will call the original braid an {\it elementary
braid} and the new one the {\it composite braid}.
When we are dealing with a link it is possible to 
replace different components by a different number 
of parallel copies. In order to handle that case
we have to consider 
mixed composite braids 
which we describe below. 

{\bf Definition}: Let ${\bf c} = (c_1,c_2, \ldots , c_n).$
An {\bf c}-composite braid of an elementary braid is
obtained by replacing the $i$-th strand by $c_i$-strands
and the
generator $b_i$ by $B_i^{({\bf c})}$ 
\eqn\mcom{B_i^{({\bf c})}= 
b (\sum_{j=1}^i c_j,\sum_{j=1}^{i+1}c_j -1 ) 
b (\sum_{j=1}^i c_j-1, \sum_{j=1}^{i+1}c_j -2 )
\dots b (\sum_{j=1}^i c_j - c_i +1, \sum_{j=1}^{i+1}c_j - c_i ). }
Clearly, for  $c_1=c_2= \dots = c_n=c$, $B_i^{({\bf c})}$ is the same as
$B_i^{(c)}$ \com.

We shall again take the Hopf link from Fig.1 as an example and 
work out the details for the $(2,3)$-cable. 
Let us 
denote the resulting diagram as $(2,3)*D_H$ and the corresponding field
theory invariant in vertical framing as $P_{\{2j_1\},\{2j_2\}}[(2,3)*D_H]$.
{\it This notation implies that $j_1$ is the representation of $SU(2)$ placed
on all the elementary strands constituting the $r_1=2$ bunch   
of strands and $j_2$ on all the elementary
strands in the $r_2=3$ bunch.} 

We are interested in determining
the invariant $P_{\{1\},\{1\}} [(2,3)*D_H]$ so that using \bre,
we get the bracket polynomial 
$\langle (2,3)*D_H \rangle$.

\centerline{\epsfxsize=3in \epsfbox{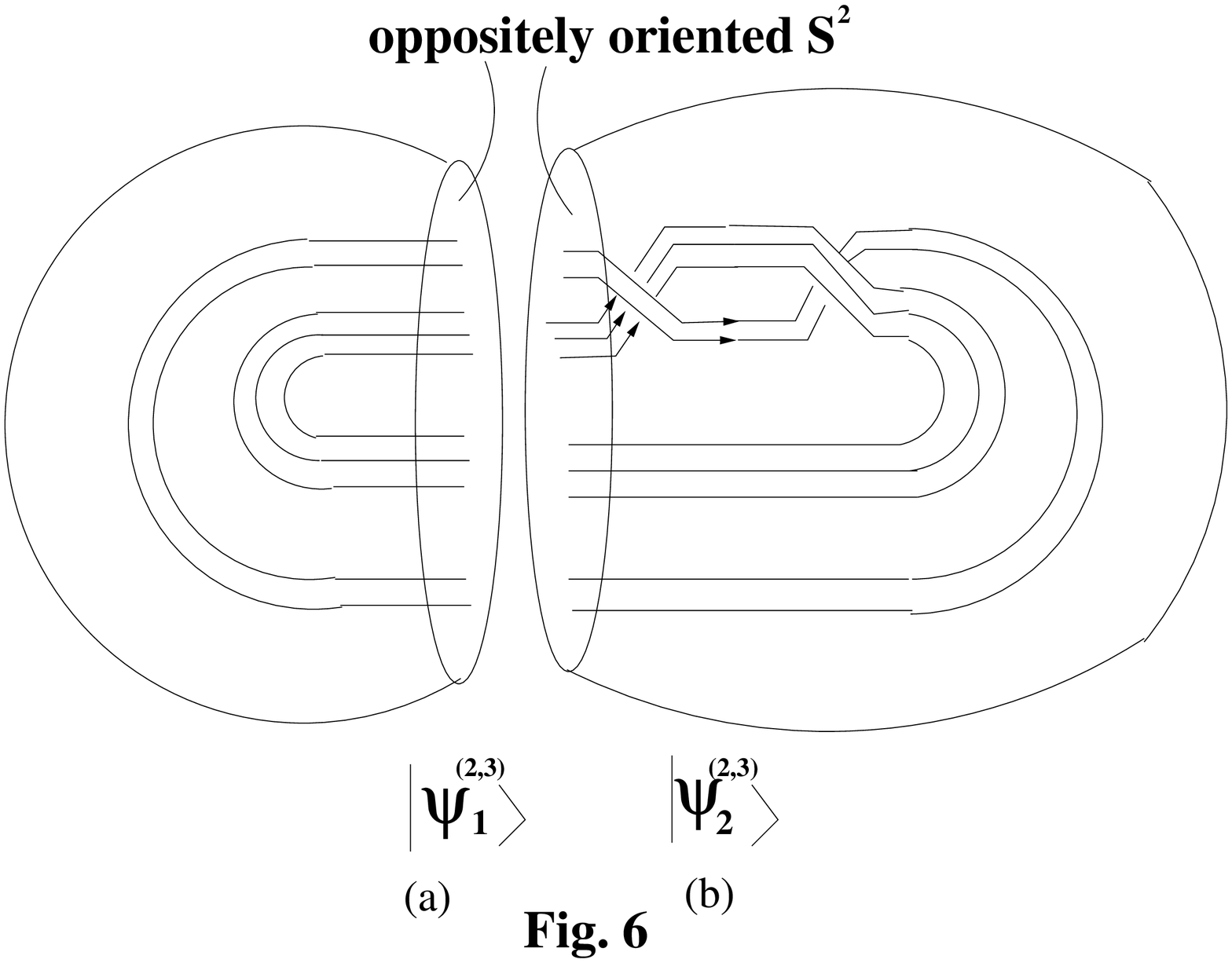}}

We present the  steps, analogous to the ones in  Section 3,
to evaluate $P_{\{1 \},\{1 \}} [ (2,3) *D_H ]$ which is obtained
by gluing the two three-balls as shown in Fig. 6(a) and (b).
Clearly, the boundary is a ten-punctured surface as
against the earlier elementary case considered in Section 3.
So, the functional integrals on these three-balls corresponds
to states in the space of ten-point correlator conformal
blocks in the Wess-Zumino conformal field theory. The
basis is so chosen that it is the eigen basis of the composite
braiding operator $B_1^{(2,3)}$. 

Using the elementary braiding
eigenvalues \pevalue, duality matrix \dual, and some of the
properties of the duality matrix which are given in
Appendix I of \kcmp, it can be shown that the eigen basis
for $(2,3)$-mixed braiding in the  side strands
is $|\phi_{(l_1,(n_1,l_2),m), ( (n_2,l_3),l_4,m)}^{side}\rangle$
with eigenvalue $\lambda_{m,R}^{(+)}(l_1,l_2)$ \pevalue\-(i.e.),
\eqn\cegn{B_1^{(3,2)}. B_1^{(2,3)} 
|\phi_{(l_1,(n_1,l_2),m), ( (n_2,l_3),l_4,m)}^{side}\rangle 
=[\hat{\lambda}_{m,R}^{(+)} (l_1, l_2) ]^2
|\phi_{(l_1,(n_1,l_2),m), ( (n_2,l_3),l_4,m)}^{side}\rangle} 
{\it The derivation of composite basis states and 
eigenvalues is along a similar direction as
elaborated in Appendix of \rom\ for a different $2$-composite braiding.}
 
Similarly, for composite braiding $(3,3)$ in the middle two strands,
we choose the basis
$|\phi_{l_1,( (n_1,l_2), (n_2,l_3), n), l_4}^{cent}\rangle$.
These basis states are pictorially depicted in
Fig. 7(a) and (b). 

\centerline{\epsfxsize=5in \epsfbox{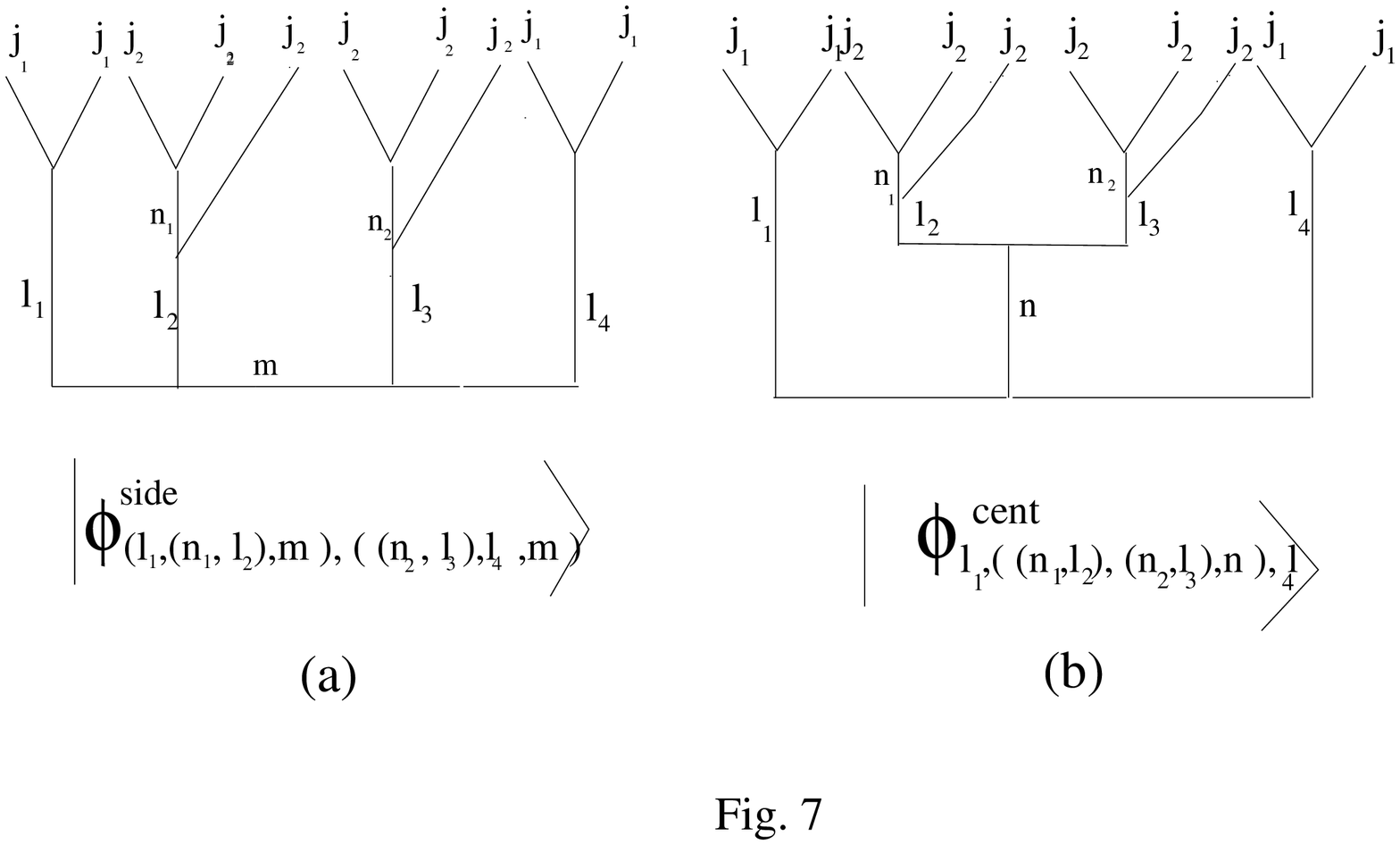}}

For $j_1=j_2={1\over  2}$,
$$l_1,n_1,n_2,l_4 \in ({1 \over 2} \otimes {1 \over 2})~;  
l_2 \in (n_1 \otimes {1 \over 2}) , l_3 \in (n_2  \otimes {1 \over 2})~;
m \in {\rm min}(l_1 \otimes l_1, l_2 \otimes l_2); 
~n \in (l_1 \otimes l_2).$$

The two bases in Fig. 7(a) and (b) are related by the
duality matrix 
\eqn\dila{
|\phi_{(l_1,(n_1,l_2),m), ( (n_2,l_3),l_4,m)}^{side}\rangle 
 = a_{mn}\left(\matrix{l_1 &
l_2 \cr l_3 & l_4 \cr}\right)
|\phi_{l_1,( (n_1,l_2), (n_2,l_3), n), l_4}^{cent}\rangle}

With these ingredients, it is clear that the state for 
Fig. 6(a) is:
\eqn\stati{|\Psi_1^{(2,3)} \rangle= 
\sum_{l_2 \in n_1 \otimes {1 \over 2}} \sum_{l_1,n_1} 
\sum_{m=|l_1-l_2|}^{{\rm min}(l_1+l_2,k-l_1-l_2)} \sqrt{[2m+1]}
|\phi_{(l_1,(n_1,l_2),m), ( (n_1,l_2),l_1,m)}^{side}\rangle~.}
The restriction $l_1=l_4$, $n_1=n_2$, $l_2=l_3$ 
in the basis states is obtained as a consequence  
of closure 
of the braid to obtain the link. The coefficients in the linear combination
are obtained from the fact that $\langle \Psi_1^{(2,3)}|\Psi_1^{(2,3)} \rangle=
P_{1,1,1,1,1}[D_{U^5}]= [2]^5$ where $D_{U^5}$ is the unlink with
five components.

Using \stati, the  state corresponding to Fig. 6(b) will be:
\eqn\staii{
\eqalign{|\Psi_2^{(2,3)} \rangle=& B_1^{(3,2)}.
B_1^{(2,3)} |\Psi_1^{(2,3)}\rangle\cr
&=\sum_{l_1,l_2,n_1}
\sum_{m=|l_1-l_2|}^{{\rm min}(l_1+l_2, k-l_1-l_2)} \sqrt{[2m+1]} 
[\hat{\lambda}_{m,R}^{(+)} (l_1,l_2)]^2 
|\phi_{(l_1,(n_1,l_2),m), ( (n_1,l_2),l_1,m)}^{side}\rangle~.}}
The summation over $n_1$ can be suppressed but we should remember
that $l_2 \in ({1 \over 2} \otimes {1 \over 2} \otimes {1\over 2})$.
Now that we have determined the states for Fig. 6(a) and (b),
the invariant for the $(2,3)$-cable of Hopf link  
can be rewritten in terms of elementary
Hopf link invariants \poly: 
\eqn\reci{P_{\{1 \},\{1 \}}[(2,3)*D_H ]=
\langle \Psi_1^{(2,3)}|\Psi_2^{(2,3)} \rangle=
\sum_{l_1,l_2} P_{2l_1,2l_2} [H]~,}
where $l_1= 0,1$ and $l_2= 1/2, 1/2, 3/2$ for $k \geq 3$.
The explicit form of the polynomial is: 
\eqn\reco{\eqalign{P_{\{1 \},\{1 \}}[(2,3)*D_H ]= & 2 P_{0,1}[D_H] + 
P_{0,3}[D_H] + 2 P_{2,1}[D_H] + P_{2,3}[D_H]~ \cr
~=~& 2 (q^{1 \over 2} + q^{-1 \over 2})
+ (q^{3 \over 2} + q^{1 \over 2} + q^{-1 \over 2} + q^{-3 \over 2}) \cr 
~&+ 2 \left( q^2(q^{1 \over 2} + q^{-1 \over 2})+ q^{-1} 
 (q^{3 \over 2} + q^{1 \over 2} + q^{-1 \over 2} + q^{-3 \over 2}) \right)\cr
~& +\left( q^5 (q^{1 \over 2} + q^{-1 \over 2}) + q^2 
 (q^{3 \over 2} + q^{1 \over 2} + q^{-1 \over 2} + q^{-3 \over 2})\right. \cr
 ~&+ \left.q^{-3} (q^{5 \over 2} + q^{3 \over 2} + q^{1 \over 2} + 
q^{-1 \over 2} + q^{-3 \over 2} + q^{-5 \over 2})\right)
\cr
~&=q^{11 \over 2}+ q^{9 \over 2} + q^{7 \over 2}+ 3 q^{5 \over 2}+
4 q^{3 \over 2}+ 6 q^{1 \over 2}\cr
&~~~~~6 q^{-1 \over 2}+ 4 q^{-3 \over 2} +
3 q^{-5 \over 2}+ q^{-7 \over 2} + q^{-9 \over 2}+ q^{-11 \over 2}
}}
The following relation is easy to check by computing
the bracket polynomial using the recursive method :
\eqn\recotwo{
P_{\{1 \},\{1 \}}[(2,3)*D_H ]\vert_{\left(q^{1 \over 4}=-A \right)}
= - \langle (2,3)*D_H \rangle ~.}
This is expected from Theorem 1 for the five component link.

Now we generalise the technique used
here for cables of arbitrary link diagrams.
A bold face lower case letter will indicate an
$n$-tuple of numbers, for ex., ${\bf c} = (c_1,c_2, \ldots , c_n)$.
Using Theorem 1 we obtain:
\vskip5mm

{\bf Theorem 2}: {\it 
The bracket polynomial of a {\bf c}-cable of the diagram of an
$n$-component link can be
expressed in terms of elementary link invariants in vertical
framing \nlink; the exact relation being:
\eqn\cbli{\langle {\bf c} 
* D_{L_n} \rangle =
(-1)^{c_1+c_2+\ldots ,c_n}
\left( \sum_{l_1,l_2 \ldots ,l_n} P_{2l_1, 2l_2, \ldots ,2l_n}[D_{L_n}] \right)
\vert_{\left(q^{1 \over 4}=-A \right)}}
where $l_i$ takes values in 
\eqn\tens{
l_i \in \underbrace{
\left( \{({1 \over 2} \otimes {1 \over 2}) \otimes {1 \over 2} \}
\ldots {1 \over 2} \right)}_{c_i}.
}}

{\bf Proof}:
Consider an {\bf c}-cable 
of an $n$-component link $L$. Suppose that $L$ is 
obtained from closure of an elementary $m$-braid. 
Let us replace 
each strand corresponding to the $i$-th component of the link
by $c_i$ parallel strands, $1 \le i \le n$.
This gives a mixed composite braid 
say $(c'_1,c'_2 \ldots ,c'_{2m})$ where as a set 
$\{ c'_1,c'_2 \ldots ,c'_{2m} \} $ is the same as 
$ \{ c_1,c_2 \ldots ,c_{n} \}$.
We will place defining representation $j=1/2$ 
on all the elementary strands. 
Closure of composite braids forces,
$$\left(c'_{m+1} \ldots ,c'_{2m} \right) = 
\left( c'_m,c'_{m-1} 
\ldots ,c'_1
\right).$$
\centerline{\epsfxsize=2in \epsfbox{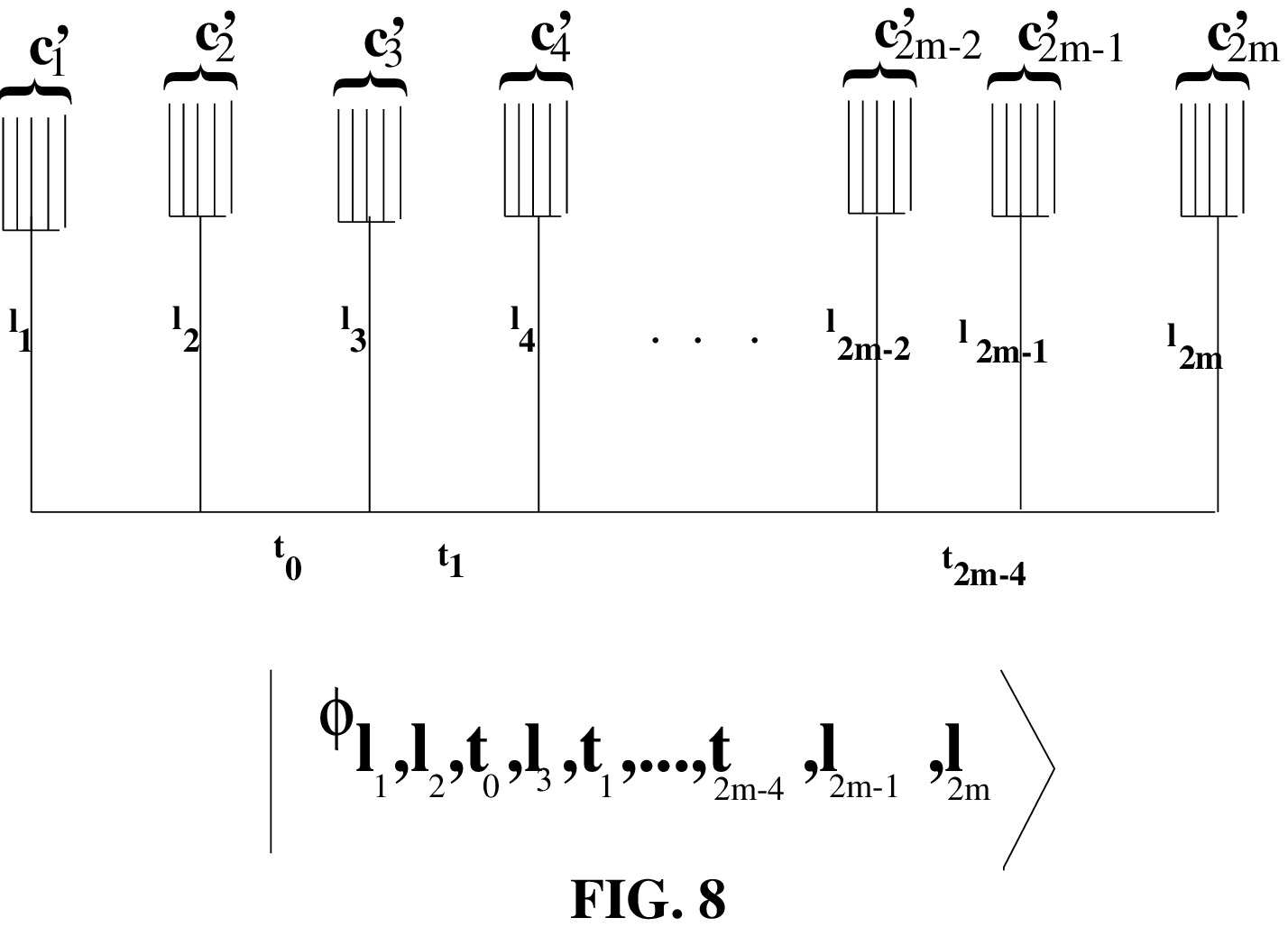}} 

Again writing the states for the mixed composite braiding in a suitable
basis (See Fig. 8):
\eqn\cstat{\eqalign{|\Psi_1^{(c_1,c_2, \ldots ,c_n)} \rangle =&
\sum_{l_1,l_2, \ldots ,l_{2m}}
\sum_{t_0,t_1, \ldots ,t_{2m-4}} A_{l_1 \ldots l_{2m},t_0 ,\ldots ,t_{2m-4}}[P]~
|{\tilde \phi}_{l_1,l_2,t_0,l_3,t_1,\ldots ,t_{2m-4},l_{2m-1},l_{2m}}
\rangle\cr
\langle \Psi_2^{(c_1,c_2, \ldots ,c_n)}| =&
\sum_{l_1,l_2, \ldots ,l_{2m}}
\sum_{t_0,t_1, \ldots ,t_{2m-4}} B_{l_1, \ldots ,l_{2m},t_0,\ldots ,t_{2m-4}}~
\langle
{\tilde \phi}_{l_1,l_2,t_0,l_3,t_1,\ldots, t_{2m-4},l_{2m-1},l_{2m}}|}}
where $t_i \in (t_{i-1} \otimes l_{i-2})$ with $t_{-1}=l_2$,
$t_{2m-3}=l_{2m}$ and $l_i$'s as in \tens.
The closure of the braid demands:
$$\left(l_{m+1}, \ldots ,l_{2m} \right) = {\cal P} \left (l_1,l_2, \ldots ,l_m
\right).$$
The constraint of closing the braid to give an 
$n$-component link, $n \leq m$ requires that 
at most $n$ of the $l_i$'s 
are distinct, and the rest are determined by the link under consideration.

Incorporating the above mentioned conditions and also using
\nlink, we get the composite invariant to be:
\eqn\conli{P_{\{1\},\{1\},\ldots,\{1\}}[
{\bf c}*D_{L_n}]
=\langle \Psi_2^{(c_1,c_2, \ldots ,c_n)}|\Psi_1^{(c_1,c_2, \ldots ,c_n)}\rangle
=\sum_{l_1,l_2,\ldots,l_n} P_{2l_1,2l_2,\ldots,2l_n
}[D_{L_n}].}
It is evident that the number of components in the
{\bf c}-cable of the $n$-component
link 
is $c_1+c_2+\ldots c_n$.
Hence Theorem 1 gives the result \cbli\ ({\bf Proved}).
\vskip5mm

We expand all the tensor products in \tens\ using \dsu\
to get some useful relations for proving the equality
of the two apparently distinct three-manifold invariants
defined by Lickorish and Kaul.
The closed form expression for the tensor product 
for $c_i \leq k$ turns out to be
\eqn\actu{l_i \in {\bigoplus_s}_{0\leq 2s \leq c_i} a_s j_s - 
{\bigoplus_t}_{c_i+2 \leq 2t \leq 2(c_i-1) }
a_t {\hat j}_t}
where representations $j_s$ and ${\hat j}_t$ are
given by $2j_s= c_i-2s$ and $2{\hat j}_t=2t-c_i-2$ 
and $a_s=\left(\matrix{ c_i-1 \cr s } \right)$ are the constants.

We use this tensor product expansion to rewrite
the bracket of an {\bf c}-cable 
in the following Corollary.
\vskip5mm

{\bf Corollary 1:} {\it With $q^{1 \over 4}=-A,$
the bracket polynomial of the ${\bf c}$-cable of an $n$-component link 
diagram $D$, where ${\bf c} = (c_1, c_2, \ldots , c_n)$, is given by
\eqn\corbrac{
        \langle {\bf c} * D \rangle~=~
               (-1)^{c_1 +c_2+ \ldots + c_n} \cdot 
\left\{ 
  \sum_{(s_1, s_2, \ldots ,s_n)}
       \left( \Pi_{i=1}^n A_{c_i,s_i} \right) P_{s_1,s_2, \ldots ,s_n}[D] 
\right\}}
where the 
$\{s_i\}$ are subjected to 
$$ {r_i - s_i\ {\rm even}~,~ 0 \leq s_i \leq c_i}, $$ and
\eqn\coef{
A_{c_i,s_i} = \left\{ 
\matrix{
\left(\matrix{c_i-1 \cr {{c_i-s_i}\over 2}}\right)
      - \left(\matrix{c_i-1 \cr {{c_i-s_i}\over 2}-2}\right),\  &
{\rm if\ } 0 \leq s_i \leq c_i-4, \cr
\left(\matrix{c_i-1 \cr {{c_i-s_i}\over 2}}\right),\  &
{\rm if\ } c_i-3 \leq s_i \leq c_i. 
} \right.
}}
\vskip5mm

The inverse of Corollary 1 expressing elementary invariants  
$P_{l_1, l_2, \ldots ,l_n}[D]$ in terms of the 
composite invariants 
$\langle {\bf c} * D \rangle$ will be useful in 
proving the equivalence between Lickorish and Kaul's
three-manifold invariants.
Hence inverting \corbrac\ we obtain the following:

\vskip5mm

{\bf Theorem 3:} {\it
For a diagram $D$ of an $n$-component link,
the invariant $P_{l_1,l_2,\ldots ,l_n}[D]$ can be expressed 
in terms of bracket polynomials as follows.
\eqn\conv{P_{l_1,l_2, \ldots ,l_n}[D] ={\sum_{\bf j}}_{0 \leq 2j_i \leq l_i}
\left\{ \Pi _{i=1}^n   (-1)^{l_i-j_i} 
\left(\matrix{l_i-j_i \cr j_i } \right)
\right\}
\langle ({\bf l} - 2{\bf j}) * D \rangle.}
}

{\bf Proof}:
We use 
Corollary 1 
to rewrite the RHS of the above equation. This changes \conv\ to:
\eqn\prfi{
P_{l_1,l_2, \ldots ,l_n}[D] ={\sum_{\bf j}}_{0 \leq 2j_i \leq l_i}
\sum_{\bf s} 
 \Pi _{i=1}^n  \left\{ (-1)^{j_i} 
\left(\matrix{l_i-j_i \cr j_i } \right)
 A_{l_i-2j_i,s_i}\right\} P_{s_1,s_2, \ldots ,s_n}[D] ,
}
where 
${\bf s}= (s_1, s_2, \ldots ,s_n)$ is such that
$l_i-s_i$ is 
even and 
$0 \leq s_i \leq l_i-2j_i$.
Let us make a change of variable $l_i-s_i= 2s_i'$,
write each sum as a multiple sum,
and interchange the summations to rewrite the statement we
need to prove as :
\eqn\frfi{\eqalign{
P_{l_1,l_2, \ldots ,l_n}[D] =& \cr
&{\sum_{s_n',j_n}}_{0 \leq  2s_n' \leq l_n, 0 \leq j_n \leq s_n'}
\ldots 
{\sum_{s_1',j_1}}_{0 \leq  2s_1' \leq l_1, 0 \leq j_1 \leq s_1'}
\cr
& \left\{ \left[ \Pi _{i=1}^n  \left( (-1)^{j_i} 
\left(\matrix{l_i-j_i \cr j_i } \right) 
 A_{l_i-2j_i,l_i-2s_i'}\right) \right] 
 P_{l_1-2s_1',l_2-2s_2', \ldots , l_n-2s_n'}[D] \right\}.
}}
We will work with these sums one at a time.
Let $$S(s_1')=
\sum_{j_1=0}^{s_1'} (-1)^{j_1}
\left(\matrix{l_1-j_1 \cr j_1 } \right)
A_{l_1-2j_1,l_1-2s_1'}
 P_{l_1-2s_1',\ldots , l_n -2s_n'}[D]~.$$
Then the first sum in \frfi\ equals
$$\sum_{s_1'=0}^{\lfloor {l_1 / 2} \rfloor} S(s_1'),$$
where $\lfloor {l_1 / 2} \rfloor$ denotes the greatest integer
less than or equal to ${l_1 / 2}$.
We split this into two sums 
$\sum_{s_1'=0}^{{\rm min}\{\lfloor l_1 / 2\rfloor ,1 \}}$ and ${\sum_{s_1'=2}^{\lfloor l_1 /2\rfloor }}$
and use eqn.\coef\ to substitute for the $A_{l_i-2j_i,s_i}$. 
It is easy to see that: 
\eqn\sumone{
{\sum_{s_1'=0}^{{\rm min}\{{\lfloor l_1 / 2\rfloor}, 1\}}} 
S(s_1')
= P_{l_1, l_2-2s_2', \ldots , l_n-2s_n'} [D].
}
For $2 \leq s_1' \leq l_1 /2$, using
\coef\ we have :
\eqn\sumtwo{\eqalign{
S(s_1')~=~&
\sum_{j_1=s_1'-1}^{s_1'} (-1)^{j_1}
\left(\matrix{l_1-j_1 \cr j_1 } \right)
\left(\matrix{l_1-2j_1-1 \cr s_1'-j_1} \right)
 P_{l_1-2s_1',\ldots , l_n -2s_n'}[D]~
\cr
&+
 (l_1-2s_1'+1) \times \cr
& \sum_{j_1=0}^{s_1'-2} 
\left[  (-1)^{j_1}
{(l_1-j_1) (l_1-j_1-1) \ldots ,(l_1-j_1-s_1'+2) \over j_1! (s_1'-j_1)! }
 P_{l_1-2s_1',\ldots ,l_n-2s_n'}[D]
\right]
.}}
We claim that this equals zero. The theorem will follow by treating
the rest of the sums in eqn. \frfi\ similarly.

In order to prove the claim first note that
$$\sum_{j_1=0}^{s_1'} {1 \over j_1! (s_1'-j_1)!} (-1)^{j_1} x^{j_1} = 
{(1-x)^{s_1'} \over s_1'!}.$$ 
Using this we see : 
\eqn\simi{\eqalign{\left( {\sum_{j_1=0}^{ s_1'}}
  (-1)^{j_1} 
{(l_1-j_1) (l_1-j_1-1) \ldots (l_1-j_1-s_1'+2) \over j_1! (s_1'-j_1)! } 
\right)\cr
~~~~~~={\rm Lt}_{x \rightarrow 1}~ (-1)^{s_1'-1} {d^{s_1'-1} \over dx^{s_1'-1}} 
\left[ \left(\sum_{j_1=0}^{s_1'} {(-1)^{j_1} x^{j_1}\over j_1! (s_1'-j_1)!}
\right) 
x^{s_1'-l_1 -2}\right]\cr
~~~~~~~={\rm Lt}_{x \rightarrow 1}~ (-1)^{s_1'-1} {d^{s_1'-1} \over dx^{s_1'-1}} 
\left\{(1-x)^{s_1'} x^{s_1'-l_1-2}\right\}~=~0 . }}
It follows that
\eqn\newsumtwo{\eqalign{S(s_1')~=&
\sum_{j_1=s_1'-1}^{s_1'} 
\left[  (-1)^{j_1} 
\left\{
\left(\matrix{l_1-j_1 \cr j_1 } \right)
\left(\matrix{l_1-2j_1-1 \cr s_1'-j_1} \right) 
\right.  \right. ~~ - \cr
&  
\left. 
(l_1-2s_1'+1) \times
{(l_1-j_1) (l_1-j_1-1) \ldots ,(l_1-j_1-s_1'+2) \over j_1! (s_1'-j)! }
\right\} \cr
&  \left. P_{l_1-2s_1',\ldots , l_n -2s_n'}[D]
\right] 
}}
A simple arithmetic shows that the expression in the RHS of 
\newsumtwo\ is 0.
%
%
{\bf (Proved)}.
\vskip5mm

Now that we have given a direct method of determining the
bracket polynomials of 
cables of link diagrams, we will 
show, 
in the next section, that 
the three-manifold invariants 
obtained from regular isotopy field theoretic invariants \two\
are the same as \one\ for the polynomial variables satisfying
eqn. \zer. 

\newsec{Conclusions}

We have given a field theoretic presentation for bracket polynomials
in terms of framed link invariants in $SU(2)$ Chern-Simons theory 
with the polynomial variable obeying \zer.
Then, using representation theory of composite braids, we 
obtained a direct method of evaluating bracket polynomials 
of cables of link diagrams. 
This enables us to show
that the three-manifold invariant obtained by
Lickorish 
using the formalisation of
the bracket polynomial as Temperley-Lieb algebra and the invariant 
obtained by Kaul using generalised link invariants from Chern-Simons theory
are equal upto normalisation.
However the normalising factor depends on the choice of $A$, the 
$4r$-th root of unity.

In the discussion below $\left(\matrix{a \cr - \cr b} \right)$ 
denotes the qudratic 
symbol for relatively prime integers $a$ and $b$,
defined as \lang\ :
\eqn\quadsym{
\left(\matrix{a \cr - \cr b} \right) ~=
\left\{ \matrix{
          +1,& {\rm if\ } a \equiv x^2\ {\rm mod\ } b,\ {\rm for\ some\ 
integer\ }x, \cr
          -1,& {\rm otherwise.}} 
\right.}

\vskip5mm

{\bf Theorem 4 :}{\it  Let $A= e^{{n \pi i} \over {2r}},$ where $n$ is a positive
integer relatively prime to $4r$ with $r$ related to the 
coupling constant in field theory as $k=r-2$.
Lickorish's invariant $F_l$ obtained from the formalisation of
the bracket polynomial as Temperley-Lieb algebra and Kaul's invariant $F_k$
obtained using generalised link invariants from Chern-Simons theory,
for the polynomial variables obeying \zer, are related as:
\eqn\final{F_k (M)~=~ \epsilon 
\kappa ^{- \nu \over 2} F_l
(M), {\ \rm where\ }}
\eqn\eps{\epsilon~=~
\left(\matrix{r \cr - \cr n } \right) e^{{(n-1)(r+1) \pi i} \over 2}
~=~ \pm 1.}}

{\bf Proof}:
It is an easy exercise in algebraic number theory (see \lang)
to show that the Gauss (qudratic) sum $G=G(e^{{n \pi i} \over {2r}})$ 
used in the definitions \likap\ and \lilam\ is
as given below.
\eqn\gauss{G= 2 \sqrt{2r} 
\left(\matrix{r \cr - \cr n } \right)
e^{{n \pi i} \over 4}.}
Clearly, ${\bar G \over G} = e^{-{n \pi i} \over 2}.$
Simplifying \likap\ and using \twoa\ it is easy to see that
\eqn\newkapb{\kappa =  \alpha^{-2}.}

Similarly using \lilam\ and \twoa\ we see that
\eqn\eight{\lambda_l= \left(\matrix{r \cr - \cr n } \right)
e^{{(n-1)(r+1) \pi i} \over 2}
\alpha^{-1}
{\sum_j}_{0\leq 2j \leq r-2-l} (-1)^{l+j} \left(\matrix{l+j \cr j} \right)
\mu_{l+2j}.}
We use Theorem 3 to write $F_k$ in terms of brackets
of cables of the diagram $D$ which represents the framed link
associated with the manifold and 
compare the coefficients of 
$\langle {\bf c} * D \rangle.$ 
We see that 
$$F_k (M)~=~
{\epsilon}^n \kappa ^{- \nu \over 2} F_l
(M).$$
Note that $\epsilon = \pm 1,$ and $n$ is odd. 
The result follows. 
({\bf Proved.})

\vskip5mm

It was shown \lo\ that with 
$A=-e^{i \pi \over 2r}$ 
Lickorish's invariant equals 
the Reshitikin Tureav invariant upto 
normalisation and a change of variable. So
from Theorem 4, it follows that Kaul's invariant defined using
the generalised link invariants in vertical framing is a reformulation of 
the Reshetikin-Turaev invariant.
The Kauffman-Lins invariant defined in 
Chapter 12 of \kaufl\ gives another normalization of the 
Witten-Reshetikin-Turaev invariant following Lickorish's Temperley-Lieb
algebra approach.

We shall rewrite the inferred result in the  following corollary.

\vskip5mm

{\bf Corollary 2}: 
{\it The  relationship
between the Kauffman-Lins invariant $Z(M)$,
which is Witten-Reshetikin-Turaev invariant upto 
a normalisation, 
and Kaul's invariant $F_k$ is 
\eqn\kaulins{
F_k (M) ~=~ Z(M) / Z(S^3),\ {\rm where\ } Z(S^3) = \mu _0
~=~ \sqrt{2 \over r} \sin \left( {\pi \over r} \right) .}}

We have shown by an indirect procedure that Kaul's
three-manifold invariant equals Witten's partition function.
It would be very interesting to see whether there is 
a direct method of deducing the above result.

\vskip5mm

{\bf Acknowledgments}: 
We would like to thank Ashoke Sen for his valuable suggestions. 
We are also grateful to T.R.~Govindarajan, R.K.~Kaul,
C.~Livingston, and J.~Prajapat for their comments.

\listrefs
\end